\documentclass[prr,twocolumn,showpacs,superscriptaddress]{revtex4-1}
\usepackage{bm,graphicx,amsmath,amssymb,color,placeins,tikz,pgflibraryarrows,braket,colortbl,xcolor,mathtools}
\usepackage[caption=false]{subfig}
\usetikzlibrary{decorations.pathreplacing}
\def\D{\partial}
\def\k{\kappa}
\def\w{\omega}
\def\a{\alpha}
\def\<{\langle}
\def\>{\rangle}
\def\ve{\varepsilon}

\begin{document}
\newcommand{\MZ}[1]{\textcolor{red}{\fbox{MZ} {\sl#1}}}

\title{Temperature dependence of the \\ optical properties of silicon
  nanocrystals}

\author{Marios Zacharias} 
\email{marios.zacharias@cut.ac.cy}
\affiliation{Research Unit for Nanostructured Materials Systems,
  Cyprus University of Technology, P.O. Box 50329, 3603 Limassol,
  Cyprus}
\affiliation{Department of Mechanical and Materials Science
  Engineering, Cyprus University of Technology, P.O. Box 50329, 3603
  Limassol, Cyprus}

\author{Pantelis C. Kelires}
\affiliation{Research Unit for Nanostructured Materials Systems,
  Cyprus University of Technology, P.O. Box 50329, 3603 Limassol,
  Cyprus}
\affiliation{Department of Mechanical and Materials Science
  Engineering, Cyprus University of Technology, P.O. Box 50329, 3603
  Limassol, Cyprus}

\date{\today}

\begin{abstract}
Silicon nanocrystals (SiNCs) have been under active investigation in the last decades and have been considered 
as a promising candidate for many optoelectronic applications including highly-efficient solar cells.
Some of the fundamental properties of interest in these nanostructures is the temperature dependence of their
optical absorption onset, and how this is controlled by different passivation regimes. 
In the present work we employ first-principles calculations in conjunction with the special displacement method 
to study the temperature dependence of the band gap renormalization of free-standing hydrogen-terminated, 
and oxidized SiNCs, as well as matrix-embedded SiNCs in amorphous silica, and we obtain good 
agreement with experimental photoluminescence data. We also provide strong evidence that the 
electron-phonon interplay at the surface of the nanocrystal is suppressed 
by oxidation and the surrounding amorphous matrix. For the matrix-embedded SiNCs, we show a high correlation
between the temperature dependence of the band gap and the Si-Si strained bonds. This result emphasizes the 
immanent relationship of electron-phonon coupling and thermal structural distortions. We also demonstrate that, 
apart from quantum confinement, Si-Si strained bonds are the major cause of zero-phonon 
quasidirect transitions in matrix-embedded SiNCs. As a final point, we clarify that, unlike optical absorption 
in bulk Si, phonon-assisted electronic transitions play a secondary role in SiNCs.

\end{abstract}


\maketitle

\section{Introduction} \label{Intro}
Over the last decades, quantum confined semiconductors based on silicon have drawn great 
scientific attention owing to their unique electronic and optical properties. 
In this regard, silicon nanocrystals (SiNCs) have enabled interesting technological applications, 
including optoelectronic devices~\cite{Pavesi_2000,Cheng_2011,Wang_2018,Zhenyi_2019}, 
quantum dot sensors~\cite{Gonzalez_2016,Kehrle_2018}, photodetectors~\cite{Neyfeh_2005,Kyun_2009,Martuza_2018} 
and bioimaging devices~\cite{McVey_2014,McVey_2018}.
Despite the numerous investigations in SiNCs, either free-standing or matrix-embedded, 
our understanding on the temperature dependence of their absorption onset and band gap renormalization
remains incomplete. This topic is of fundamental and practical interest for 
optimizing the efficiency of next generation solar
cells~\cite{Conibeer_2008,Luo_2011,Shuangyi_2016,Mazzarella_2019,Mazzaro_2019}, 
and silicon-based photonics~\cite{Yuan_2009,Valenta_2019}.

Various temperature-dependent photoluminescence (PL) measurements of SiNCs
have been reported~\cite{Heitmann_2004,Wang_2005,Rinnert_2009,Hartel_2012,Kusova_2012}. 
It has been found that the energies of PL peaks of matrix-embedded SiNCs 
in amorphous silica (a-SiO$_2$) exhibit a Varshni behavior~\cite{Varshni_1967}, 
following closely the band gap renormalization of bulk Si.
However, the origin of this observation is still unclear.
K$\mathring{\rm u}$sov\'a {\it et al}.~[\onlinecite{Kusova_2012}] have reported 
temperature-dependent PL spectra of free-standing oxide-passivated SiNCs.
Their results reveal that, as the temperature increases, the PL energy shift of 
free-standing SiNCs is significantly larger than the corresponding shift of matrix-embedded SiNCs. 
This difference has been attributed solely to compressive strain exerted on the nanocrystals 
by the a-SiO$_2$ matrix, ignoring the effect of atomic vibrations on the electronic structure.

The interaction of electrons with quantized atomic vibrations, namely the electron-phonon coupling,
is the underlying mechanism associated with temperature-dependent optical 
properties of materials~\cite{FG_review}. In particular, electron-phonon coupling is responsible, 
among others, for the quantum zero-point renormalization and temperature dependence 
of the electronic energy levels. This latter effect determines the thermally induced 
energy shifts of the absorption onset in bulk and nanostructured semiconductors.

In this study we rely on a recently developed methodology~\cite{Zacharias_2016,Zacharias_2020}, 
namely the special displacement method (SDM), and present first-principles calculations of 
temperature-dependent band gaps of SiNCs. We demonstrate that the electron-phonon coupling leads to a larger 
band gap renormalization in free-standing than matrix-embedded SiNCs. 
To explain this difference we also calculate Eliashberg spectral functions and analyze 
the contribution to the band gap renormalization from individual phonon modes. 
Importantly, our results reveal that surface oxidation and embedding strongly suppress the coupling 
of electrons with phonons at the surface of the SiNC. Our findings also confirm that the band gap renormalization 
of the SiNC/a-SiO$_2$ system is in good agreement with the Varshni-like temperature-dependence of bulk Si.

In addition to modifying the band structure, the electron-phonon coupling plays an important
role in optical absorption leading to phonon-assisted electronic transitions. 
For example, in indirect gap crystals, optical transitions between the band extrema require 
the participation of a phonon to satisfy the momentum conservation rule. 
In nanocrystals made of indirect gap solids, however, the intensity of the absorption onset is 
also determined by zero-phonon quasidirect recombination that breaks the momentum conservation rule. 
This effect has been found to be pronounced in SiNCs, especially in 
relatively small size clusters~\cite{Hybertsen_1994,Kovalev_1998,Benjamin_2016}. 

Quantum confinement (QC) is the primary factor controlling quasidirect transitions. However, 
the absorption onset of SiNC structures is found to be consistently well below the quantum confined 
band gap~\cite{Luppi_2005}. Various suggestions have been made in this respect. These are based
on oxygen-related surface/interface states~\cite{Wolkin_1999,Luppi_2005,Hadjisavvas_2007,Benjamin_2012}, 
chemical passivation by other ligands~\cite{Dohnalov_2014}, strained Si-Si bonds at the interface layer
\cite{Allan_1996,Hadjisavvas_2004,Benjamin_2012}, and interface scattering~\cite{Benjamin_2016}.

In this manuscript, we also report results of first-principles calculations that 
unambiguously identify some of the main sources of quasidirect transitions in the
SiNC/a-SiO$_2$ system, and clarify the role of each of its components
(NC core, interface, matrix). This is made possible by the decomposition
of density of states and optical absorption into atomic contributions, probing 
in this way the optical response of individual sites and local areas in the system.
Our results show that, apart from QC, the dominant source of quasidirect transitions 
originates from strained Si-Si bonds in the core region induced by embedding.

The organization of the manuscript is as follows: in Sec.~\ref{sec.Theory}
we briefly introduce the theory and main equations employed to investigate 
temperature-dependent optical properties of SiNCs, and outline the recipe used 
to obtain site decomposition of optical absorption. 
Section~\ref{sec.Methods} reports all computational details of the calculations 
performed in this work. In Sec.~\ref{sec.Results} we present our results on
various SiNC systems. In particular, in Sec.~\ref{sec.Results_Size_SiNCs}
we report band gaps of H-terminated, oxidized, and matrix-embedded SiNCs 
as a function of the nanocrystal diameter and compare to experiment.
In Sec.~\ref{sec.Results_Tgap_SiNCs} we compare our calculations of 
temperature-dependent band gaps of H-terminated, oxidized, and matrix-embedded SiNCs 
(all with diameter 2~nm) with bulk Si and experimental data. 
These calculations are followed by the analysis of phonon density of states 
and Eliashberg spectral functions of the three SiNC systems in Sec.~\ref{sec.Results_Eliashberg}.
In Sec.~\ref{sec.Results_Role_Si-Si_strain} we present the role 
of Si-Si strained bonds in the optical properties of matrix-embedded SiNCs, and 
in Sec.~\ref{sec.Results_Quasidirect}, for the same structure, we compare Tauc plots of zero-phonon and 
phonon-assisted spectra. In Sec.~\ref{Sec.Conclusions} we summarize our key findings 
and indicate avenues for future work.

\section{Theory} \label{sec.Theory}

In this section, we briefly describe the theoretical framework of the
methodology employed to calculate temperature-dependent optical properties,
and the decomposition of the dielectric function into site contributions. 
In the following, we adopt a $\Gamma$-point formalism, since our calculations 
are for free-standing and extended matrix-embedded SiNCs. 

\subsection{Temperature-dependent optical properties} \label{sec.Theory_1}

Within the Williams-Lax~\cite{Williams_1951,Lax_1952} theory in the harmonic approximation, the imaginary part of the
dielectric function at photon frequency $\w$ and temperature $T$ is given by~\cite{Zacharias_2015}: 
 \begin{equation}\label{eq.eps_HA_T}
  \epsilon_2(\w,T) = {\prod}_\nu \int\! dx_\nu \frac{\exp(-x_\nu^2/2\sigma_{\nu,T}^2)}{\sqrt{2\pi}\sigma_{\nu,T}}
  \epsilon_2^{\{x_\nu\}}(\w).
  \end{equation}
Here the multi-dimensional Gaussian integral is taken over all normal coordinates $x_\nu$ and the superscript 
denotes the dielectric function evaluated with the nuclei in configuration $\{x_\nu\}$. The widths of the 
Gaussian distributions are defined by the mean square displacements of the atoms along a vibrational mode $\nu$, 
with frequency $\w_\nu$, and is given by $\sigma^2_{\nu,T}  = (n_{\nu,T}+1/2) \, \hbar/ M_{\rm p} \w_{\nu}$, 
where $M_{\rm p}$ is the proton mass and $n_{\nu,T} = [\exp(\hbar\w_\nu/k_{\rm B}T)\!-\!1]^{-1}$ 
represents the Bose-Einstein occupation. 
The fundamental approach to numerically evaluate Eq.~\eqref{eq.eps_HA_T}
is to employ Monte Carlo integration and perform an average of the dielectric functions
calculated for multiple atomic configurations $\{x_\nu\}$. Those configurations are constructed independently 
by generating normal coordinates from the multivariate normal distribution 
${\prod}_\nu \exp(-x_\nu^2/2\sigma_{\nu,T}^2)/\sqrt{2\pi}\sigma_{\nu,T}$. 

Recently, it has been shown that one can identify a single atomic 
configuration~\cite{Zacharias_2016,Zacharias_2020}, 
namely the ZG configuration, to evaluate the integral in Eq.~\eqref{eq.eps_HA_T}. 
The set of special atomic displacements that lead to the ZG configuration are obtained via:    
\begin{equation}\label{eq.ZG_displ}
   \Delta \tau_{\k\a} = \sqrt{\frac{M_p}{M_\k}}\sum_\nu (-1)^{\nu -1} \,  e^\nu_{\k\a} \, \sigma_{\nu,T}, 
\end{equation}
where  $\Delta \tau_{\k\a}$ is the displacement of atom $\k$ along the Cartesian direction $\a$, 
and $e^\nu_{\k\a}$ is the component of the phonon polarization vector 
associated with the normal mode coordinate $x_{\nu} = \sigma_{\nu,T}$. 
The calculation of optical spectra with the ZG configuration yields correctly the adiabatic limit 
of phonon-assisted optical absorption as described by Hall, Bardeen, and Blatt~\cite{hbb_1954}, and at the same 
time incorporates the effect of electron-phonon renormalization on the band structure.

In the same way with the Williams-Lax optical spectra, one can calculate temperature-dependent transition energies 
by directly replacing the dielectric function in Eq.~\eqref{eq.eps_HA_T} with the 
transition energy~\cite{Zacharias_2020}. 
In the thermodynamic limit, the band gap evaluated for the ZG configuration at temperature $T$ is given, 
up to fourth order in atomic displacements, by~\cite{Zacharias_2016}: 
 \begin{eqnarray}\label{eq.energy_HA_T}
  E_{\rm g}(T) &=& E_{\rm g} + \frac{1}{2} \sum_\nu \frac{\D^2 E_{\rm g}}{\D x_{\nu}^2} \sigma^2_{\nu,T} 
 \\ &+& \frac{3}{4!} \sum_{\mu \neq \nu} \frac{\D^4 E_{\rm g}}{\D x_{\mu}^2 \D x_{\nu}^2} 
     \sigma^2_{\mu,T} \sigma^2_{\nu,T}, \nonumber
     + \frac{1}{4!} \sum_{\nu} \frac{\D^4 E_{\rm g}}{ \D x_{\nu}^4} \sigma^4_{\nu,T},
\end{eqnarray}
where $E_{\rm g}$ indicates the band gap energy evaluated with the 
nuclei of the system at their equilibrium geometry. The 
quadratic terms consist of the Fan-Migdal and Debye-Waller contributions 
to the electron-phonon renormalization~\cite{Patrick_2014,FG_review},
and the fourth order terms [last line of Eq.~\eqref{eq.energy_HA_T}] 
represent two-phonon contributions. 

A standard way to investigate the vibrational mode contribution to the band gap renormalization 
at finite temperatures is to calculate the Eliashberg spectral function, defined as:
\begin{equation}\label{eq.Eliash_sp_fn}
g^2F_{\rm g}(\w,T) = \sum_\nu \frac{1}{2}\frac{\D^2 E_{\rm g}}{\D x_{\nu}^2} \sigma^2_{\nu,T} \, \delta (\hbar\w - \hbar\w_\nu) \, .
\end{equation}
Integrating $g^2F_{\rm g}(\w,T)$ over all phonon energies yields the 
sum of the Fan-Migdal and Debye-Waller corrections. 
This sum is usually evaluated in state-of-the-art perturbative 
calculations~\cite{Giustino_2010,Cannuccia_2011,Gonze_2011,Ponce_2014,Ponce_2014_2,Antonius_2014,Villegas_2016,Lihm_2020} 
of temperature-dependent band structures that rely on the Allen-Heine theory~\cite{Allen_1976}.

\subsection{Site decomposition of optical spectra} \label{sec.Theory_2}

The imaginary part of the dielectric function in the independent particle and electric dipole approximations
is given by~\cite{Cardona_Book}: 
\begin{equation}\label{eq.diele_fn}
  \epsilon_2(\w) = 2\frac{4 \pi^2 e^2}{m_e^2 \w^2 V} \sum_{v c} 
        | p_{cv}|^2 \delta(\ve_c - \ve_v -\hbar \w). \ 
\end{equation}
In this expression $m_e$ and $e$ are the electron mass and charge, the factor of two is for the spin degeneracy, 
$V$ is the volume of the system, and $p_{cv} = \<\psi_c|\nabla|\psi_v\>$ is the optical matrix element representing
direct transitions between the valence and conduction Kohn-Sham 
states $|\psi_v\>$ and $|\psi_c\>$ with energies $\ve_v$ and $\ve_c$, respectively.

Taking the expansion of the Kohn-Sham states as a linear combination of atomic orbitals, the 
optical matrix element is written as:
\begin{equation}\label{eq.opt_mat_ele_LCAO}
        p_{cv} = \sum_{ij} c^*_{i c} c_{j  v}  \<\phi_{i}|\nabla|\phi_{j}\> 
               = \sum_{ij} c^*_{i c} c_{j  v} {\tilde p}_{ij}, 
\end{equation}
where $c_{j v}$ represents the Kohn-Sham expansion coefficients of the state $|\psi_v\>$,  and
$\phi_{j}$ are the corresponding atomic basis states. To alleviate the notation we also define 
the matrix elements describing transitions between the basis states as 
${\tilde p}_{ij} =  \<\phi_{i}|\nabla|\phi_{j}\>$.

Now we split the system into two groups of atoms, $A$ and $B$, so that the total optical matrix element 
can be decomposed into the following contributions: 
\begin{eqnarray}\label{eq.opt_mat_ele_LCAO_decomp}
p_{cv} = p^{AA}_{cv} + p^{BB}_{cv} + p^{AB}_{cv} + p^{BA}_{cv}, 
\end{eqnarray}
where
\begin{eqnarray}\label{eq.opt_mat_ele_LCAO_decomp_2}
p^{AB}_{cv} &=& \sum_{i\in A,j\in B} c^*_{i c} c_{j  v} {\tilde p}_{ij}. 
\end{eqnarray}
Here the self matrix elements, $p^{AA}_{cv}$ and $p^{BB}_{cv}$, represent optical transitions between 
basis states that are associated entirely with atoms in the same group, while the 
cross-coupling matrix elements, $p^{AB}_{cv}$ and $p^{BA}_{cv}$, represent optical transitions between 
basis states that are associated with atoms in different groups. 

By substituting Eq.~\eqref{eq.opt_mat_ele_LCAO_decomp} into Eq.~\eqref{eq.opt_mat_ele_LCAO} and expanding 
the square modulus we can obtain the decomposition of the dielectric function
into site contributions. The result is:
\begin{eqnarray}\label{eq.epsilon2_decomp}
\epsilon_2(\w) = \epsilon^{AA}_2(\w) + \epsilon^{BB}_2(\w) + {\rm cct}
\end{eqnarray} 
where, for example, the dielectric function 
\begin{eqnarray}\label{eq.epsilon2_decomp_2}
\epsilon^{AA}_2(\w) &=& 2\frac{4 \pi^2 e^2}{m_e^2 \w^2 V} \sum_{v c} 
        |p^{AA}_{cv}|^2 \delta(\ve_c - \ve_v -\hbar \w)
\end{eqnarray} 
corresponds to optical absorption from atoms in group $A$ only, and ``${\rm cct}$'' refers to 
the various cross coupling terms. We note that the above prescription to decompose optical absorption 
into site contributions can be generalized straightforwardly to an arbitrary number of groups.

\section{Computational details and Methods} \label{sec.Methods}
All first-principles calculations are based 
on density functional theory using numeric atom-centered orbitals as basis functions
in the PBE generalized gradient approximation~\cite{GGA_Pedrew_1996} as implemented  
in the electronic structure package FHI-aims~\cite{Blum_2009}.
The sampling of the Brillouin zone of each nanocrystal was performed using 
the $\Gamma$ point, and a vacuum of at least 20 \AA \, was considered in all Cartesian directions
to avoid spurious interactions between periodic images. Ground state geometries were obtained via 
BFGS optimization~\cite{Nocedal_2006} until the residual force component per atom was 
less than 10$^{-2}$ eV/\AA. 

\begin{figure}[t!]
\includegraphics[width=0.45\textwidth]{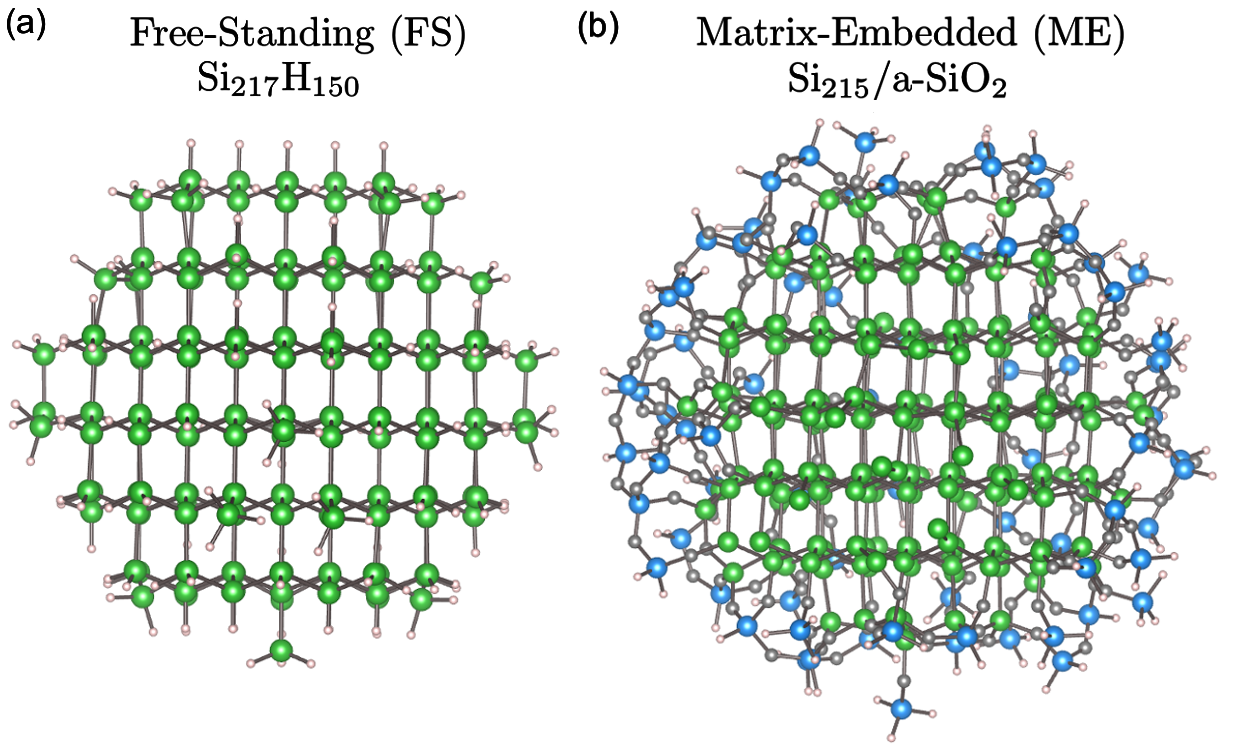}
\caption{\label{fig1} Structures of (a) H-terminated free-standing (FS) and (b) matrix-embedded 
(ME) SiNCs with diameter $d=2$~nm. 
Si atoms in the nanocrystals and passivating H atoms are shown in green and white. Matrix Si and O atoms are
colored blue and gray, respectively.
} 
\end{figure}

The dangling bonds of free-standing SiNCs are passivated with 
hydrogen atoms, as shown in Fig.~\ref{fig1}(a). Oxidized free-standing 
SiNCs were obtained by replacing two hydrogens with one oxygen atom followed by
geometry optimization. Configurations containing only double-bonded oxygens were considered.
Unlike Si-O-Si bridge bonds, Si$=\hspace{-0.15cm}=$O double 
bonds are known to vary significantly the electronic structure of relatively small SiNCs~\cite{Luppi_2005}.
For example, our calculations on free-standing SiNCs with $d = 1.2$ nm (Si$_{45}$H$_{58}$) 
reveal that the formation of Si$=\hspace{-0.15cm}=$O and Si-O-Si bonds 
cause a band gap reduction of 845~meV and 214~meV, respectively. 
 

Initial matrix-embedded SiNCs were obtained by a well tested Monte Carlo (MC)
approach, as described in detail elsewhere \cite{Hadjisavvas_2004,Hadjisavvas_2007}. 
They consist of small spherical NCs with a size of 1.4 nm (99 Si atoms) and 2.0 nm 
(215 Si atoms), embedded in a-SiO$_2$ matrices containing 400 and 2450 oxide atoms, respectively.
To make the first-principles calculations tractable in the latter case, we constructed from the original 
network a smaller cell containing 659 atoms, in which the large NC is coated with an oxide layer 
of only 292 atoms, representing the rest of the oxide, and passivated with 152 H
atoms, as shown in Fig.~\ref{fig1}(b). These initial structures were first thoroughly annealed and
relaxed with {\it ab initio} Molecular Dynamics at 800~K. Then, they were cooled to 0~K and brought
into their ground state via geometry relaxation. The final structures 
contain only Si-O-Si bridge bonds, which have been shown to form the lowest 
energy configuration at the interface~\cite{Hadjisavvas_2004}.
 We note that no Si$=\hspace{-0.15cm}=$O double bonds could be stabilized 
under relaxation, even after artificially imposing them, which indicates their 
high formation energy in embedded NCs\cite{Luppi_2007,Pennycook_2010}.

All calculations of temperature-dependent optical properties were performed 
on SiNCs with diameter $d = 2$~nm.
The structures with the nuclei at their relax positions are shown in Fig~\ref{fig1}.
The Williams-Lax theory in the harmonic approximation together with 
SDM~\cite{Zacharias_2016,Zacharias_2020}[Eq.~\eqref{eq.ZG_displ}] 
were employed to calculate dielectric functions and band gaps 
at finite temperatures. Optical matrix elements were calculated within the independent 
particle approximation by taking the isotropic average over the Cartesian directions.
Vibrational frequencies $\w_\nu$ and eigenmodes $e^\nu_{\k\a}$ were obtained via the frozen-phonon 
method~\cite{Kunc_Martin,Ackland} as implemented in PHONOPY~\cite{phonopy}. 
ZG configurations were generated via Eq.~\eqref{eq.ZG_displ}. 
For the ZG displacement in the H-passivated matrix-embedded SiNC we excluded 
the modes associated with displacements of H atoms. This choice 
avoids to a large extent the spurious contributions to the band gap 
renormalization coming from the artificial passivation to the system. All calculations 
of temperature-dependent band gaps and spectra were performed using the ZG configuration and 
its antithetic pair~\cite{Zacharias_2016} to ensure that contributions from 
the linear terms in atomic displacements reduce to zero. The derivatives ${\D^2 E_{\rm g}}/{\D x_{\nu}^2}$ required to 
obtain the Eliashberg spectral function in Eq.~\eqref{eq.Eliash_sp_fn} were calculated by finite differences~\cite{Capaz_2005}.
This procedure required $2\times3P$ frozen-phonon calculations, where $P$ is the number of atoms 
in the system. 

To elucidate the origin of the phonon density of states and band gap renormalization 
in different structures we decompose the atoms of the nanocrystal into two groups: surface and core atoms. 
The core contains Si atoms having as neighbors only Si$^{0}$, 
i.e. Si atoms not bonded to any O or H atoms. 
The surface of free-standing 
SiNCs contains H, O, and passivated Si atoms (suboxides Si$^{+1}$, Si$^{+2}$, and Si$^{+3}$).
The surface of the matrix-embedded SiNC contains matrix 
atoms (suboxide Si$^{+4}$ and O), and passivated Si atoms (suboxides Si$^{+1}$, Si$^{+2}$, and Si$^{+3}$).

To account for the effect of thermal expansion on the band gap of SiNCs we performed the following steps:
(i) We took the volume of all nanocrystals to be equal to the convex hull volume formed by the outermost Si atoms. 
(ii) We mimicked volume expansion by increasing the distance 
of all Si atoms from the center of the nanocrystal by the same fraction. 
That is, we increased the volume without inducing distortion to the bond angles,
and thus keeping the structures similar. 
(iii) For the case of matrix-embedded SiNCs, we neglected the thermal expansion of the matrix 
and left the positions of amorphous silica atoms at their relaxed coordinates.  
The rationale behind this choice is that the thermal expansion coefficient of amorphous silica~\cite{Lager_1982}
(0.5$\times$10$^{−6}$ K$^{-1}$) is five times less than that of bulk Si~\cite{Tokumaru_1984}
(2.6$\times~10^{−6}$ K$^{-1}$). (iv) We took the thermal expansion coefficient of SiNCs to be 2.5 
times larger than that of bulk Si~\cite{Freitas_2012}.

\section{Results} \label{sec.Results}

\subsection{Size dependent band gap of SiNCs} \label{sec.Results_Size_SiNCs}

\begin{figure}[t!]
\includegraphics[width=0.48\textwidth]{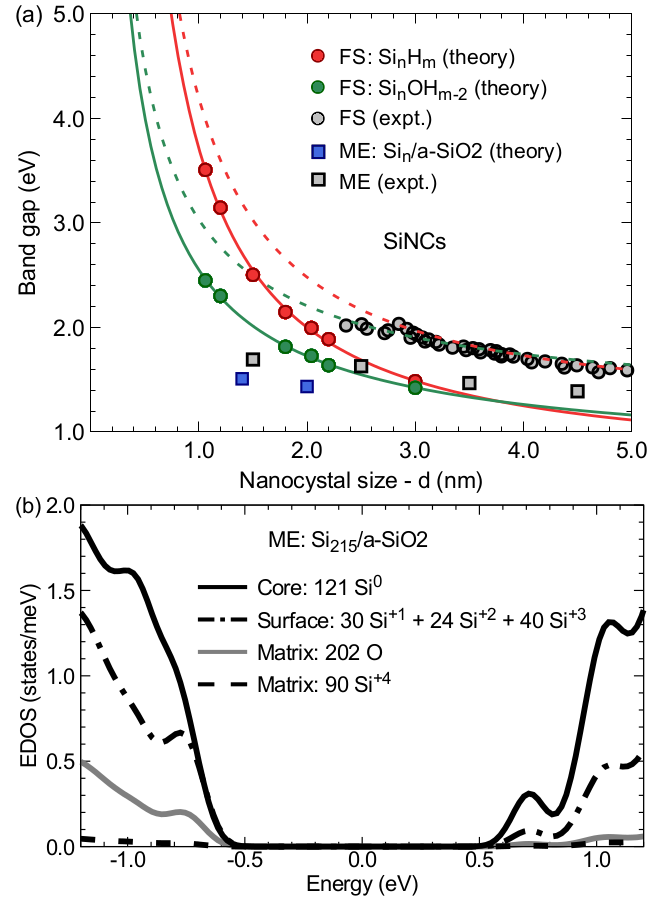}
\caption{\label{fig2}
  (a) Band gap of silicon nanocrystals (SiNCs) as a function 
  of average diameter $d$. Calculated band gaps of H-terminated (Si$_n$H$_m$), 
  oxidized SiNCs (Si$_n$OH$_m$) and matrix-embedded (Si$_{n}$/a-SiO$_2$) SiNCs 
  are shown as red discs, green discs, and blue squares, respectively. 
  The solid curves are fits of the form
  $E_0+a/d^{b}$ to the calculated band gaps of H-terminated SiNCs (red) 
  and oxidized SiNCs (green)~\cite{Delerue_1993}.
  The dashed curves represent vertically shifted theoretical fits 
  such that $E_0$ corresponds to the GW corrected band gap of bulk Si~\cite{Lebegue_2003}.  
  Experimental data of matrix-embedded SiNCs in a-SiO$_2$ (black squares) 
  and free-standing SiNCs with oxidized surface (black discs) are from 
  Refs.~[\onlinecite{Hartel_2012}] and~[\onlinecite{Ledoux_2002}].
  (b) Decomposition of the electronic density of states (EDOS) of the matrix-embedded SiNC 
  into contributions from Si core atoms (black solid line), passivated surface Si atoms (black dashed dotted line),
  O matrix atoms (gray solid line), and Si matrix atoms (black dashed line).
  The Fermi level is set at the middle of the gap. A Gaussian broadening of 70 meV was used.
}
\end{figure}

Figure~\ref{fig2}(a) shows the calculated band gaps of H-terminated free-standing SiNCs/Si$_n$H$_m$ (red discs),
oxidized free-standing SiNCs/Si$_n$OH$_{m-2}$ (green discs), and two embedded SiNCs in a-SiO$_2$ (blue squares)
as a function of the nanocrystal diameter $d$. 
Experimental data of embedded SiNCs in a-SiO$_2$ (black squares)~\cite{Hartel_2012} 
and free-standing SiNCs with oxidized surface (black discs)~\cite{Ledoux_2002} are shown for comparison. 
Our results show that the band gap of free-standing SiNCs opens with decreasing nanocrystal size.
As expected, this trend follows the QC theory and compares favorably 
with other electronic structure calculations~\cite{Proot_1992,Delerue_1993,Wang_Zunger_1997,Puzder_2002}.
The significant reduction of oxidized and matrix-embedded SiNCs from the QC values is also clear.
For example, our calculations on nanocrystals with $d=2$~nm reveal a band gap reduction of 0.27~eV and 0.43~eV 
after oxidation and embedding, respectively. 
The latter value is in close agreement with 0.4~eV extracted from experimental data of SiNCs with $d=2.5$~nm.

The solid curves in Fig.~\ref{fig2}(a) represent fits to the data of 
free-standing SiNCs and are of the form $E_0 + a/d^b$, where $E_0=0.66$~eV is the 
calculated PBE band gap of bulk Si and $a$, $b$ are fitting parameters. 
The fit to Si$_n$H$_m$ band gaps gives $a=3.04$ and $b=1.19$, 
which compare nicely with the theoretical values of $3.73$ and 
$1.39$ obtained in Ref.~[\onlinecite{Delerue_1993}]. In both cases the
exponent $b$ is different from 2 showing that the effective mass model
is inadequate to describe the energy levels of nanocrystal clusters~\cite{Rama_1991,Luo_2006}. 
The fit to Si$_n$OH$_m$ band gaps gives $a=1.88$ and $b=0.83$. 
We attribute the further decrease of $a$ and $b$ to the new electronic states that appear
near the band edges as a result of the formation of a Si$=\hspace{-0.15cm}=$O double bond~\cite{Luppi_2005}.
The systematic underestimation of the measured band gaps is mainly due to 
the PBE approximation to the exchange-correlation energy used in our calculations. 
As illustrated by the dashed green and red curves in Fig.~\ref{fig2}(a), this underestimation is adjusted 
by a ``scissor'' shift equal to 0.5 eV that mimics the GW quasiparticle 
corrections to the band gap of bulk Si~\cite{Lebegue_2003}. 

Figure~\ref{fig2}(b) shows the decomposition of the electronic density of states (EDOS) 
near the band gap of the matrix-embedded SiNC with $d=2$~nm. The main contribution to the EDOS is from 
Si core and surface atoms. At variance with Si matrix atoms, O atoms participate 
in the formation of the band edges leading to the reduction of the band gap energy
from its QC value~\cite{Benjamin_2012}. 
However, this is not the primary factor contributing to the band gap closing and
thereby to quasidirect transitions close to the absorption edge; 
as we demonstrate in Secs.~\ref{sec.Results_Role_Si-Si_strain} and~\ref{sec.Results_Quasidirect} embedding causes 
large strains to Si-Si bonds that alter significantly the electronic band structure and optical absorption.

\begin{figure*}[hbt!]
\includegraphics[width=0.99\textwidth]{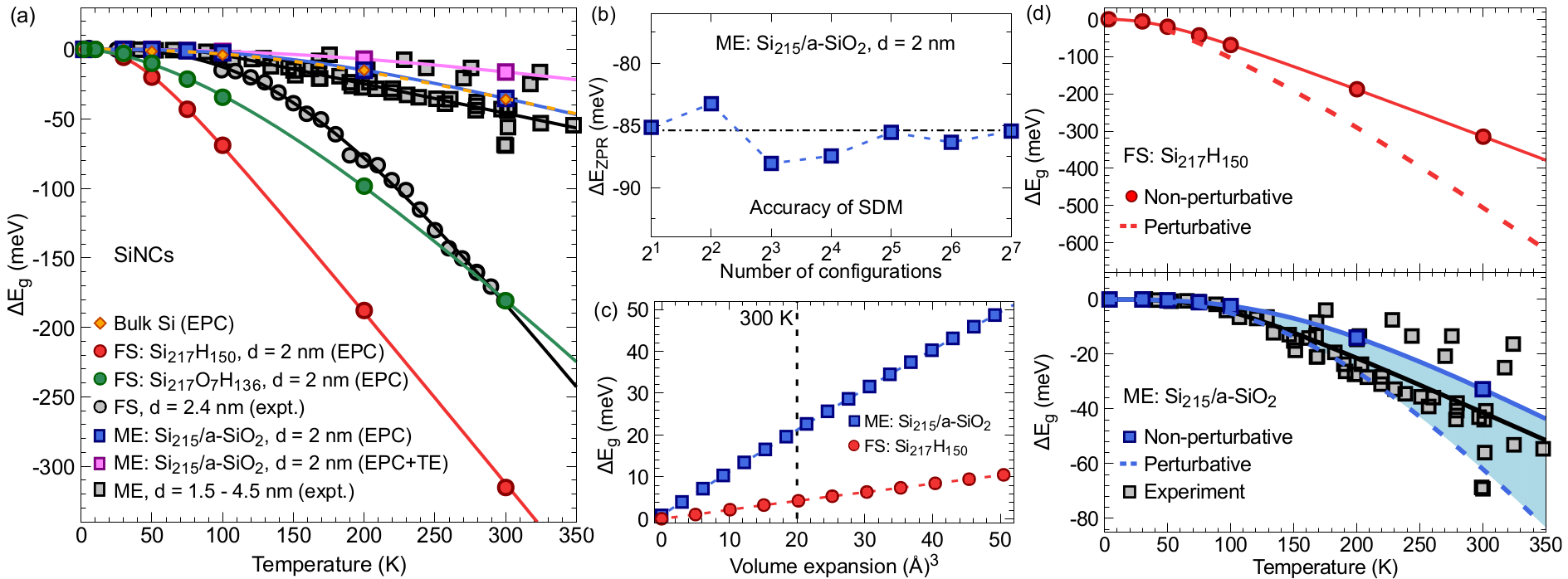}
\caption{\label{fig3}
  (a) Temperature dependence of the band gap renormalization of free-standing 
  (FS) and matrix-embedded (ME) SiNCs up to 350~K.
  Calculated band gaps using the ZG displacement~\cite{Zacharias_2016} for H-terminated (Si$_{217}$H$_{150}$), 
  oxidized (Si$_{217}$O$_7$H$_{136}$) and matrix-embedded (Si$_{215}$/a-SiO$_2$) SiNCs are shown as red discs, green discs
  and blue squares, respectively. The diameter of the nanocrystals employed for our calculations is 2 nm. 
  Magenta squares represent the band gaps of matrix-embedded SiNCs including also the effect of 
  thermal expansion (TE).
  Orange diamonds represent data of bulk Si reported in Ref.~[\onlinecite{Zacharias_2020}].
  Experimental data of oxidized free-standing SiNCs (black discs)
  is from  Ref.~[\onlinecite{Kusova_2012}] and of matrix-embedded SiNCs in a-SiO$_2$ (black squares) is from
  Refs.~[\onlinecite{Wang_2005,Rolver_2005,Rinnert_2009,Hartel_2012}].
  All curves represent fits to Eq.~\eqref{eq.Passler_oscil_model}.
  (b) Variation of the zero-point renormalization of the matrix-embedded SiNC with the number of deterministic ZG configurations. 
      The horizontal dashed line indicates the zero-point renormalization of 86~meV calculated using only one antithetic pair of ZG configurations.
  (c) Band gap renormalization versus volume expansion of H-terminated SiNCs (red discs) and 
      matrix-embedded SiNCs in a-SiO$_2$ (blue squares). 
      The vertical dashed line indicates the volume expansion at 300~K.
  (d) Comparison of temperature-dependent band gaps of the H-terminated SiNC (top panel) and 
      matrix-embedded SiNCs in a-SiO$_2$ (bottom panel) calculated using the 
      SDM (solid lines) and finite differences (dashed lines). 
      The shaded area can be taken as the uncertainty of the band gap renormalization 
      calculated for matrix-embedded SiNCs in a-SiO$_2$.
      Experimental data of matrix-embedded SiNCs is as for (a).
 }
\end{figure*}

\subsection{Temperature-dependent band gap of SiNCs} \label{sec.Results_Tgap_SiNCs}

In Fig.~\ref{fig3}(a) we compare our calculations of the temperature-dependent band gap renormalization of SiNCs  
with experiments from Refs.~[\onlinecite{Kusova_2012,Wang_2005,Rolver_2005,Rinnert_2009,Hartel_2012}], up to $T=350$~K.
To facilitate comparison with data obtained from photoluminescence measurements we define the band gap renormalization 
as $ \Delta E_{\rm g}(T) = E_{\rm g}(T) - E_{\rm g}(0)$.
The effect of electron-phonon coupling (EPC) is included in our calculations using SDM. 
Red discs, green discs and blue squares represent calculations of fully H-terminated, oxidized  and matrix-embedded 
SiNCs with $d=2$~nm, respectively. 
The oxidized SiNC (Si$_{217}$O$_7$H$_{136}$) was prepared by replacing 14 hydrogen 
with 7 oxygen atoms to form Si$=\hspace{-0.15cm}=$O 
double bonds that are uniformly distributed in the outer-shell of the nanocrystal. 

As shown in Fig.~\ref{fig3}(a),
oxidation of the nanocrystal leads to a significantly smaller variation of the band gap with temperature. 
For example, our results reveal a decrease in the band gap renormalization of 135 meV at 300~K, amounting 
to 43\% of the band gap renormalization calculated for the fully H-terminated SiNC. 
This difference is explained by the fact that the oxidized nanocrystal exhibits a weaker 
electron-phonon coupling at the surface; this aspect is analyzed in detail in Sec.~\ref{sec.Results_Eliashberg}.
Our calculated temperature-dependent band gaps of the oxidized SiNC 
compare well with measurements~\cite{Kusova_2012} made on free-standing oxide-passivated SiNCs 
with $d=2.44$~nm (black discs). We note that we did not explore how the formation of 
Si-O-Si bridge bonds, or other oxidation combinations~\cite{Luppi_2005}, affect the 
temperature dependence of the band gap. Such a study requires systematic and extended 
analysis, and will be presented elsewhere.

Figure~\ref{fig3}(a) also shows that electron-phonon renormalization of the band gap is further reduced by 
placing the nanocrystal inside the a-SiO$_2$ cage. 
The interpretation of this result is given in Sec.~\ref{sec.Results_Eliashberg}.
Our data (blue squares) exhibits a very similar behavior to the band gap renormalization 
calculated for bulk Si~\cite{Zacharias_2020} (orange diamonds).
Compared to experimental data~\cite{Wang_2005,Rolver_2005,Rinnert_2009,Hartel_2012,Kusova_2012} (black squares)
reported for nanocrystals with diameter 1.5 -- 4.5 nm the agreement is very good.
As already shown for bulk Si in Ref.~[\onlinecite{Karsai_2018}], 
this agreement can be improved by taking into account GW corrections to the electron-phonon coupling.
For completeness, we include in our calculations the additional change in the band gap 
resulting from the thermal expansion of the nanocrystal inside the matrix [magenta squares in Fig~\ref{fig3}(a)]. 
Also in this case, the theoretical data lie within the experimental range. 

\begin{table}[b!] 
 \captionsetup{font=small}
\caption{Fitting parameters $a$, $\Theta$ and $p$ entering Eq.~\eqref{eq.Passler_oscil_model} to describe the temperature dependence 
of the band gap renormalization of free-standing (FS) and matrix-embedded (ME) SiNCs, and bulk Si. 
TE indicates that the effect of thermal expansion is included. Experimental values of matrix-embedded 
SiNCs and bulk Si are from Refs.~[\onlinecite{Rinnert_2009}] and~[\onlinecite{Cardona_2001}], respectively.}
        \hspace*{-0.2cm}
\centering
        \begin{tabular*}{0.45\textwidth}{l *{4}c }
        \hline\hline \\  [-0.25 cm]      
                              &      $a$         &  $\Theta$              &  $p$  & $\Delta E_{\rm ZPR}$      \\ [0.05 cm]  
                              &  (meV K$^{-1}$)  & \,\,\,\,  (K) \,\,\,\, &  \,\, & (meV)     \\ [0.05 cm]  \hline \hline \\ [-0.62 cm]\\
        FS: Si$_{217}$H$_{150}$           & 1.29             &  118                   & 2.26  & 230  \\ [0.01 cm] 
        FS: Si$_{217}$O$_7$H$_{136}$ & 1.07             &  486                   & 1.68  & 141  \\ [0.01 cm] 
        ME: Si$_{215}$/a-SiO$_2$     &  0.25            & 365                    & 3.00  & 86   \\ [0.05 cm]
        ME with TE         &  0.15            & 534                    & 2.45  & 86   \\ [0.05 cm]
        Bulk Si               &  0.26            &  425                   & 2.47  & 57   \\ [0.05 cm]
        FS (expt.)            & 1.32             & 365                    &  2.99 &  -      \\ [0.05 cm]
        ME (expt.)            & 0.17 -- 0.4      & 68 -- 400              &  2.5 -- 2.8& -    \\ [0.05 cm]
        Bulk Si (expt.)       & 0.3176           & 406                    & 2.33  & 64   \\ [0.05 cm] \hline \hline
        \end{tabular*}
\label{table.1}
\end{table}

All curves in Fig.~\ref{fig3}(a) represent fits to the band gap renormalization 
$\Delta E_{\rm g}(T)$ of SiNCs
using an alternative expression to Varshni's law, 
 given by\cite{Passler_1996_b}: 
\begin{eqnarray} \label{eq.Passler_oscil_model}
\Delta E_{\rm g} (T) = -\frac{a \Theta}{2} 
 \Bigg[ \sqrt[\leftroot{+1}\uproot{6} {\huge p}]{1 + \bigg(\frac{2T}{\Theta}\bigg)^p} -1 \Bigg],
\end{eqnarray}
where $a$, $\Theta$ and $p$ are model parameters representing the gradient of the high-temperature asymptote, 
the effective phonon temperature and the exponent of the temperature power law. 
This expression was chosen to (i) describe more accurately the non-linear dependence of the band gap 
at very low temperatures\cite{Passler_1996}, and (ii) facilitate comparison with available experimental data\cite{Rinnert_2009}.
The values of $a$, $\Theta$ and $p$ obtained from our analysis and the corresponding experimental parameters from
Refs.~[\onlinecite{Rinnert_2009}] and~[\onlinecite{Cardona_2001}] are summarized in Table~\ref{table.1}. 

As a sanity check, we performed calculations to test the accuracy of SDM 
for the case of finite size nanocrystal clusters. Figure~\ref{fig3}(b) shows 
the change of the average zero-point renormalization (ZPR) of the band gap with the number 
of ZG displacements generated for the matrix-embedded SiNC, d = 2 nm. The zero-point renormalization of the band gap 
is defined as $\Delta E_{\rm ZPR} = E_{\rm g}(0) - E_{\rm g}$. 
The ZG displacements were generated such that the numerical error in the 
evaluation of the ZPR is reduced with configurational averaging~\cite{Zacharias_2016}.
Our results confirm that $\Delta E_{\rm ZPR} = 86$ meV is already well converged
using a single antithetic pair of ZG displacements. 
The calculated $\Delta E_{\rm ZPR}$ for all nanocrystals and bulk Si obtained using 
 SDM are listed together with experimental data of bulk Si in Table~\ref{table.1}.

Figure~\ref{fig3}(c) shows the band gap change as a function of homogeneous volume expansion 
of the free-standing H-terminated (red discs) and matrix-embedded (blue squares) SiNCs. 
The approach we employed to mimic the volume expansion of SiNCs is provided in Sec.~\ref{sec.Methods}.
The band gap of both structures increases linearly with volume. 
For the free-standing structure, the linear fit gives a slope of 0.2 meV/\AA$^3$, 
and the band gap opening is attributed solely to the expansion of the average Si-Si bond length. 
For the matrix-embedded nanocrystal, the linear fit gives a relatively much larger slope of 1 meV/\AA$^3$. 
This difference is explained by the presence of matrix oxygen atoms at the interface. 
In particular, as the nanocrystal expands inside the matrix, 
the overlap between the orbitals of oxygen and outermost silicon atoms increases.
Since oxygen orbitals contribute to the electronic density of states at the band edges [Fig.~\ref{fig2}(b)], 
then the decrease/distortion of Si-O-Si bridge bonds at the surface of the nanocrystal
 leads effectively to an additional band gap opening. 
To account for the effect of volume expansion to the 
band gap renormalization in matrix-embedded SiNCs we take the thermal expansion coefficient of the 
nanocrystal equal to $6.5\times10^{−6}$ K$^{-1}$; see for details in Sec.~\ref{sec.Methods}. 
This amounts to a volume increase of 20 \AA$^3$ at 300~K as indicated by the vertical dashed line 
in Fig.~\ref{fig3}(c). We also note that by taking the bulk modulus of the nanocrystal to be equal 
to 105 GPa~\cite{Kleovoulou_2013}, the average compressive stress exerted on the nanocrystal by 
the matrix at 300~K can be estimated to be 1.6 GPa. 
This value is in line with the corresponding average stress calculated in Ref.~[\onlinecite{Kleovoulo_2013_b}] 
using Monte Carlo simulations. 

\begin{figure}[b!]
\includegraphics[width=0.48\textwidth]{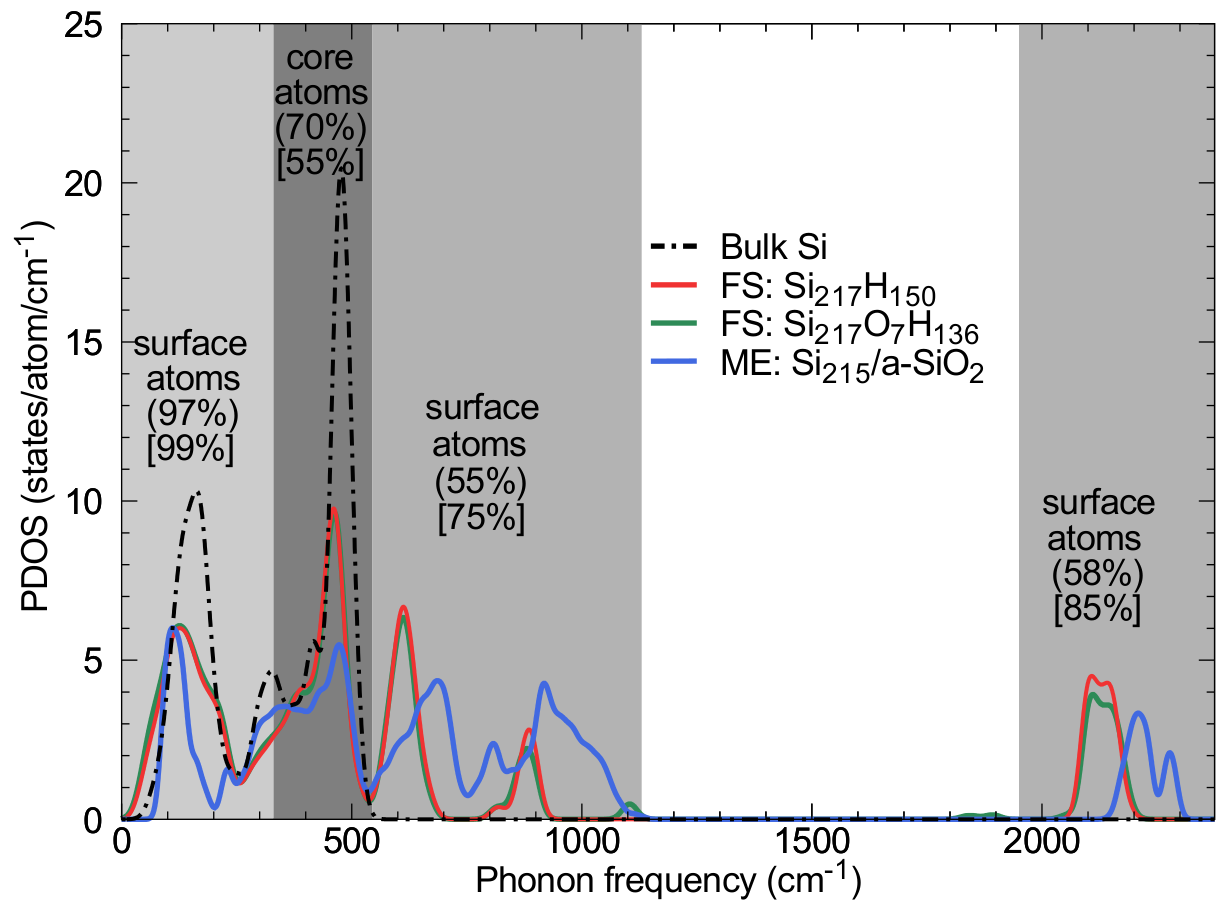}
\caption{\label{fig4}
  Phonon density of states (PDOS) of the H-terminated free-standing SiNC (red), 
  oxidized free-standing SiNC (green), matrix-embedded SiNC in a-SiO$_2$ (blue), and bulk Si (black).
  The shaded areas define the frequency range of the vibrational modes associated mainly with
  displacements of surface (light gray) and core atoms (dark gray).
  Percentages in round [square] brackets show the contribution to 
  the vibrational modes associated with displacements of the indicated group of atoms in the
  free-standing [matrix-embedded] SiNC.}
\end{figure}

Figure~\ref{fig3}(d) shows the temperature-dependent band gap renormalization of the 
H-terminated free-standing (top panel) and matrix-embedded (bottom panel) SiNCs calculated using
SDM (solid lines) and finite differences~\cite{Capaz_2005} (dashed lines). 
These methodologies are used to evaluate the temperature dependence of the band gap within 
the non-perturbative adiabatic and perturbative adiabatic Allen-Heine theory~\cite{Allen_1976,FG_review}, respectively.
The advantage of non-perturbative over perturbative approaches is that higher order electron-phonon coupling terms, 
beyond second-order perturbation theory, are included in the calculation of temperature-dependent observables 
[see for example the terms in the last line of Eq.~\eqref{eq.energy_HA_T}].
Regarding the band gap renormalization of bulk Si, these terms are known to be negligible~\cite{Zacharias_2016,Ponce_2014}. 
However, this is not the case for H-terminated free-standing SiNCs. Our finite difference calculations reveal   
$\Delta E_{\rm ZPR }= 340$ meV and a high temperature asymptote $a=2.36$ meV K$^{-1}$, 
which are well above than the corresponding values calculated using SDM 
(see Table~\ref{table.1}). We attribute this difference to the presence of light mass H atoms 
at the surface of the free-standing SiNC. In particular, vibrational modes that are associated with 
large displacements of H atoms couple to modes of the nanocrystal, or to each other, 
leading to a non-negligible higher order electron-phonon 
coupling renormalization. Further analysis of this aspect requires a separate set of elaborate calculations, 
and is beyond the scope of this manuscript.
Our perturbative calculations for the matrix-embedded SiNC give $\Delta E_{\rm ZPR }= 182$ meV and $a = 0.54$ meV K$^{-1}$,
which are more than twice the corresponding non-perturbative values (see Table~\ref{table.1}). 
The difference between our perturbative and non-perturbative calculations can 
be explained by the remaining spurious displacements of the artificially imposed H atoms that 
contribute to the band gap renormalization. This uncertainty in our calculations is shown as a blue shaded area 
in Fig.~\ref{fig3}(d).

\begin{figure}[b!]
\includegraphics[width=0.47\textwidth]{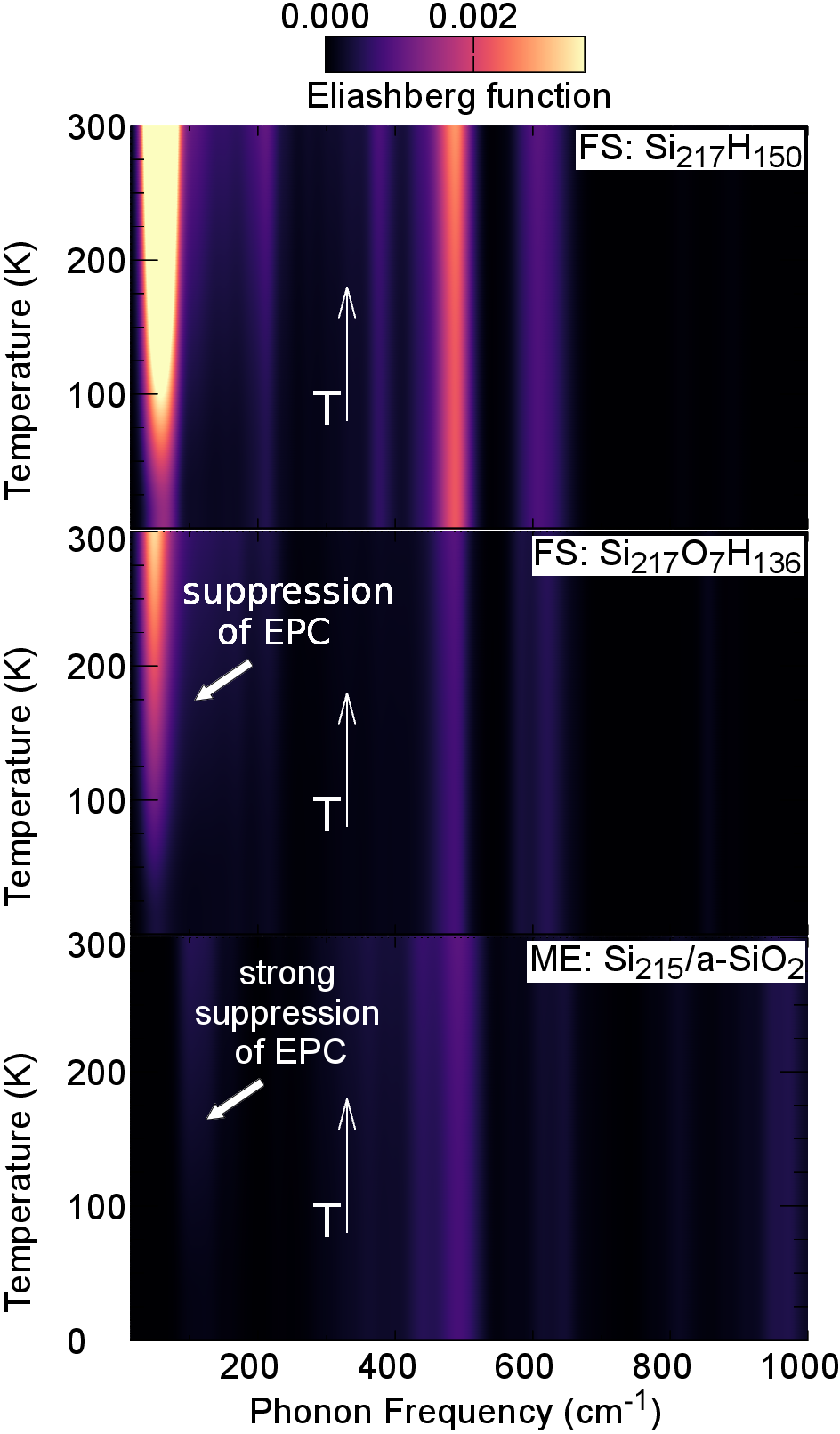}
\caption{\label{fig5}
Temperature dependence of the Eliashberg spectral function up to 300~K versus phonon frequency calculated for 
the free-standing fully H-terminated SiNC, the free-standing oxidized SiNC, and the matrix-embedded SiNC in a-SiO$_2$ 
from top to bottom, respectively. Thin arrows indicate the direction of increasing temperature and 
thick arrows highlight the suppression of electron-phonon coupling (EPC) after oxidation, 
and after placing the nanocrystal inside the matrix.} 
\end{figure}

\begin{figure*}[hbt!]
\includegraphics[width=0.97\textwidth]{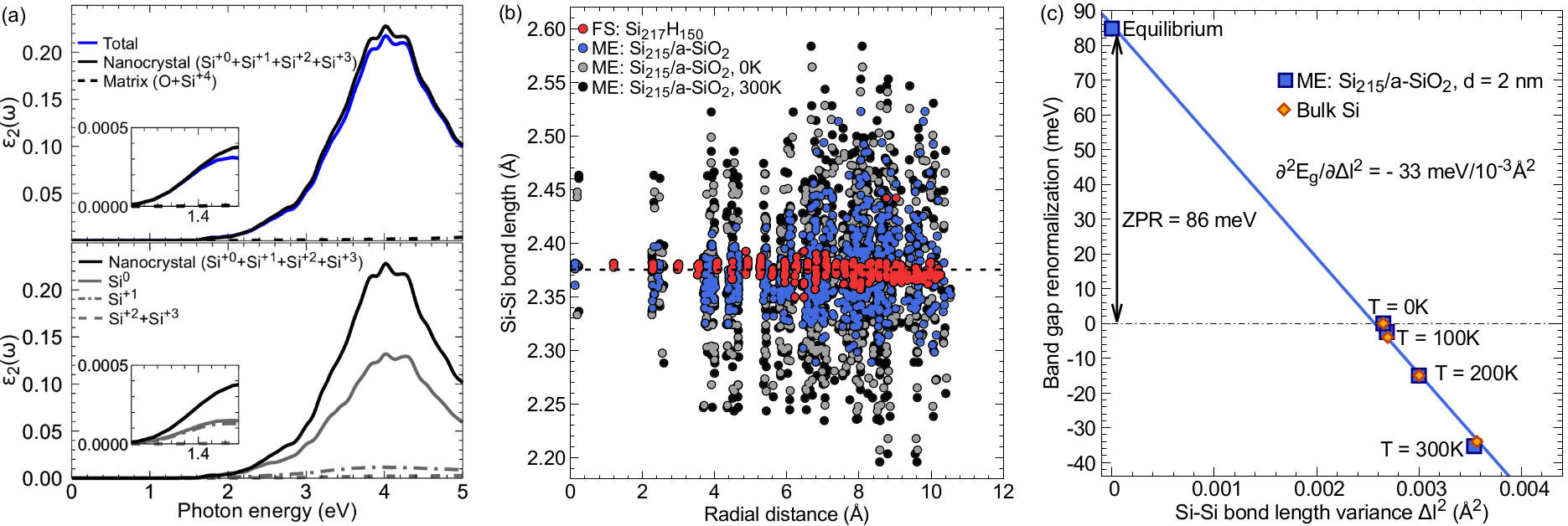}
\caption{\label{fig6}
 (a) Top panel: Decomposition of the imaginary part of the dielectric function $\epsilon_2(\omega)$ of the
     matrix-embedded SiNC (blue solid line) into site contributions from nanocrystal (black solid line) 
     and matrix (black dashed line) atoms.
     Bottom panel: Decomposition of the imaginary part of the dielectric function $\epsilon_2(\omega)$ of the 
     embedded nanocrystal (black solid line) into 
     site contributions from suboxides: Si$^0$ (gray solid line), Si$^1$ (gray dashed line) 
     and Si$^2$$+$Si$^3$ (green solid line). A Gaussian broadening of 50 meV was used in all plots.
 (b) Si-Si bond length in free-standing (red) and matrix-embedded (blue) SiNCs as a function of
     the radial distance from the center of the nanocrystal. The variation of Si-Si bond length
     in the matrix-embedded SiNC after applying ZG displacements is also shown for $T=0$~K (gray) and $T=300$~K (black). The 
     horizontal black line indicates the average Si-Si bond length (2.375 \AA) in the matrix-embedded SiNC with 
     the nuclei clamped at their equilibrium positions.
 (c) Band gap renormalization versus Si-Si bond length variance [Eq.~\eqref{eq.bond_length_var}] 
     calculated for matrix-embedded SiNCs (blue squares) that correspond 
     to equilibrium and ZG geometries for four temperatures. 
     The blue line represents the linear regression [Eq.~\eqref{eq.bond_length_var_2}] to the data with slope 
     $\partial^2 E_{\rm g} / \partial \Delta l^2 =$-33~meV/10$^{-3}$ \AA$^2$.
     Data for bulk Si (orange diamonds) are included for comparison.}
\end{figure*}

\subsection{Phonon DOS and Eliashberg spectral functions}\label{sec.Results_Eliashberg}

To explain the different behavior between the temperature-dependent band gaps
of the three SiNCs employed in our calculations, we proceed with the analysis of their phonon
density of states (PDOS) and the temperature dependence of their Eliashberg spectral functions (ESFs).

Figure~\ref{fig4} shows our calculated PDOS of the H-terminated (red curve), 
oxidized (green curve), and matrix-embedded (blue curve) SiNCs.
For comparison purposes, we also include data of bulk Si (black dashed curve).
We conclude that surface oxidation has a minor effect on the PDOS of free-standing SiNCs. 
Instead, placing the nanocrystal inside the matrix leads to distinct changes in the PDOS,
and in particular for phonon frequencies larger than 500 cm$^{-1}$, i.e. 
beyond the phonon frequency range of bulk Si.
In the same figure we also indicate the frequency ranges (shaded areas) and percentage contribution
of the vibrational modes associated mainly with displacements of surface atoms (light gray), 
and core atoms (dark gray). Details for this classification are provided in Sec.~\ref{sec.Methods}. 
  
Figure~\ref{fig5} shows the temperature dependence of the ESFs up to 300~K 
calculated for the three SiNCs in the phonon frequency range 
0 -- 1000 cm$^{-1}$. We exclude from our analysis the vibrational modes with 
frequencies higher than 1000 cm$^{-1}$, since their contribution to the 
total ZPR is less than 1\%. Furthermore, these high frequency modes have much lower occupation 
probability according to the Bose-Einstein distribution. It is evident from our calculated
ESFs that the band gap renormalization of the two free-standing SiNCs 
is dominated by the low frequency vibrational modes associated with displacements of surface atoms.
On the contrary, the largest contribution to the band gap renormalization of
the matrix-embedded SiNC comes from the modes associated with displacements of core atoms.  
It is also pronounced that oxidation, although does not cause any significant changes in the PDOS, 
it suppresses the coupling of electrons with 
surface phonons, justifying the sizable difference between the band gap renormalization 
of oxidized and fully H-terminated SiNCs, reported in Fig.~\ref{fig2}(a). 
The underlying mechanism responsible for this behavior is the participation 
of oxygen orbitals to the formation of the electron states close to band edges~\cite{Luppi_2005},
thereby modifying the associated electron-phonon coupling matrix elements~\cite{FG_review}.
Even more remarkably, the interaction of electrons with surface phonons is 
strongly suppressed when the nanocrystal is embedded in amorphous silica.  
This suggests that the large difference between the temperature-dependent
photoluminescence shifts measured for oxide-passivated and matrix-embedded SiNCs
in Ref.~[\onlinecite{Kusova_2012}] is mainly due to the strong suppression of 
electron-phonon coupling at the surface of the nanocrystal.
We attribute to the same effect, the observation that the band gap renormalization
of matrix-embedded SiNCs follows closely the Varshni behavior of bulk Si data.
In fact, as demonstrated in Ref.~[\onlinecite{Monserrat_2014}]
the dominant contribution to the electron-phonon coupling in
bulk Si originates from the phonon modes in the frequency 
range 415 - 515 cm$^{-1}$, which coincides to a large extent with the frequency range of 
the modes associated with displacements of the core atoms in the matrix-embedded SiNC.

\subsection{Role of S\lowercase{i}-S\lowercase{i} strained bonds in the optical properties 
of  matrix-embedded S\lowercase{i}NC\lowercase{s} } \label{sec.Results_Role_Si-Si_strain}

In this section, we analyze the role of Si-Si strained bonds in the 
optical absorption and band gap renormalization of matrix-embedded SiNCs.

Figure~\ref{fig6}(a) shows the spectra decomposition into groups of atoms 
of the matrix-embedded SiNC calculated with the nuclei at their equilibrium geometry.
Details of the method are available in Sec.~\ref{sec.Theory_2}.
The top panel shows the total imaginary part of 
the dielectric function  (blue solid line) and its decomposition into nanocrystal (black solid line) 
and matrix (black dashed line) self contributions. 
It is evident that the optical spectrum is dominated by the nanocrystal site,
while the matrix contribution is essentially zero. This suggests that
although matrix oxygen states participate in the formation of the absorption 
edge [Fig.~\ref{fig2}(b)], they do not participate actively in the absorption process.
The bottom panel of Fig.~\ref{fig6}(a) shows the spectra decomposition of the
nanocrystal into suboxide contributions. Our results reveal that for energies well above 
the absorption onset the largest contribution to $\epsilon_2(\w)$ originates from 
the core Si$^{0}$ atoms (gray solid line), which are not bonded to any O atoms. 
Closer to the absorption onset [inset of Fig.~\ref{fig6}(a)], the core atoms contribute
as much as the inner-interface Si$^{+1}$ atoms (gray dashed line), while 
the central Si$^{+2}$ and outer Si$^{+3}$ atoms (green solid line) are very weak absorbing elements.
The above findings lead us to the conclusion that the band gap decrease from the QC values
in matrix-embedded SiNCs is due to silicon states, mainly in the nanocrystal core,
while oxygen states play a secondary role. As we demonstrate below, this effect has 
its origins in the large bond length strains induced by embedding, previously suggested by 
Ref.~[\onlinecite{Hadjisavvas_2004}], both at the interface and deeper in the core.

Figure~\ref{fig6}(b) shows the equilibrium Si-Si bond lengths in free-standing (red discs) and matrix-embedded (blue discs) 
SiNCs, $d=2$~nm, as we move radially outwards from the center of the nanocrystal.  
The average equilibrium Si-Si bond length in the matrix-embedded SiNC is indicated by 
the horizontal black dashed line cutting the vertical axis at 2.375~\AA. This value 
differs by only 0.005~\AA\, from the corresponding value calculated for the free-standing SiNC.
It can be readily seen that the vast majority of Si-Si bond lengths in the free-standing SiNC
are very close to the average value, giving a standard deviation $\Delta l = 0.007$~\AA.
On the contrary, for the matrix-embedded SiNC, the variation of Si-Si bond lengths from the average 
is substantial, giving a standard deviation $\Delta l =  0.037$~\AA. 
In Fig.~\ref{fig6}(b), we also include the Si-Si bond lengths 
in ZG geometries of the matrix-embedded system at $T=0$~K (gray discs) and $T=300$~K (black discs). 
The corresponding standard deviations are 0.062~\AA\,and 0.068~\AA. 
These large differences in the standard deviations exemplify the large strain
induced by embedding and thermal effects, and can be correlated with the band gap 
closing in matrix-embedded SiNCs. 

To clarify this observation, Fig.~\ref{fig6}(c) shows the relationship between the 
calculated band gap renormalization of the matrix-embedded SiNC and the 
Si-Si bond length variance $\Delta l^2$. 
We define the bond length variance at temperature $T$ as: 
\begin{equation}\label{eq.bond_length_var}
 \Delta l^2 (T) = \frac{1}{S}\sum_{i=1}^S [l_i(T) - l^0_i]^2,
\end{equation}
where $l_i(T)$ and $l_i^0$ are the bond lengths in ZG and equilibrium geometries, respectively, 
and $S$ is the number of total bond lengths. 
Our choice of a quadratic deviation measure is rationalized by the fact that, 
the electron-phonon renormalization, for relatively low temperatures, is predominantly quadratic 
in atomic displacements [first line of Eq.~\eqref{eq.energy_HA_T}]. Therefore, we proceed with  
the following simple relationship
\begin{equation}\label{eq.bond_length_var_2}
 \Delta E_{\rm g}(T) = \Delta E_{\rm ZPR} + \frac{\partial^2 E_{\rm g}}{ \partial \Delta l^2}\Delta l^2 (T) ,
\end{equation}
to correlate the band gap change and bond length distortion. 
The fit of Eq.~\eqref{eq.bond_length_var_2} to the calculated band gaps 
(equilibrium and four temperatures up to 300~K) gives a slope 
$\partial^2 E_{\rm g} / \partial \Delta l^2 =$-33~meV/10$^{-3}$ \AA$^2$, 
$\Delta E_{\rm ZPR} = 86$~meV, and a Pearson's linear correlation coefficient $R=-0.9988$. 
This result confirms the linear variation of $\Delta E_{\rm g}(T)$ with $\Delta l^2$ 
emphasizing the major importance 
of Si-Si strained bonds on the electronic structure of matrix-embedded SiNCs.

In Fig.~\ref{fig6}(c) we also show that the corresponding data 
calculated for bulk Si using ZG configurations (orange diamonds)~\cite{Zacharias_2020}
almost overlap with those obtained for the matrix-embedded SiNC.
This finding further supports the conclusion that the 
band gap renormalization of the matrix-embedded SiNC 
behaves very similarly to that of its bulk counterpart.

\subsection{Quasidirect optical absorption in S\lowercase{i}NC\lowercase{s}} \label{sec.Results_Quasidirect}
In this section we analyze the Tauc plots~\cite{Tauc_1966} of SiNCs and investigate the origin of 
quasidirect optical transitions~\cite{Hybertsen_1994,Benjamin_2016} in these nanoscale structures. 
We present calculations of the imaginary part of the dielectric function 
from which the absorption properties follow directly. 

In bulk semiconductors with an indirect band gap, optical transitions between the states of the valence band maximum (VBM) and 
conduction band minimum (CBM) are forbidden in the absence of phonons owing to the
 momentum conservation rule~\cite{Cardona_Book}. 
This is indeed the case for bulk Si for which it has been shown, 
from first-principles~\cite{Noffsinger_2012,Zacharias_2015}, that phonon-assisted 
transitions are required to reveal absorption for photon energies below the fundamental direct gap.
However, it is well known that structural perturbations can relax the momentum conservation rule  
leading to zero-phonon transitions between the band edges~\cite{Hybertsen_1994,Benjamin_2016,Kusova_2014}. 
In particular, deviations from translational invariance of the crystal can cause the overlap between the electron and 
hole wavefunctions allowing for quasidirect (vertical) transitions with finite probability.

\begin{figure}[t!]
\includegraphics[width=0.50\textwidth]{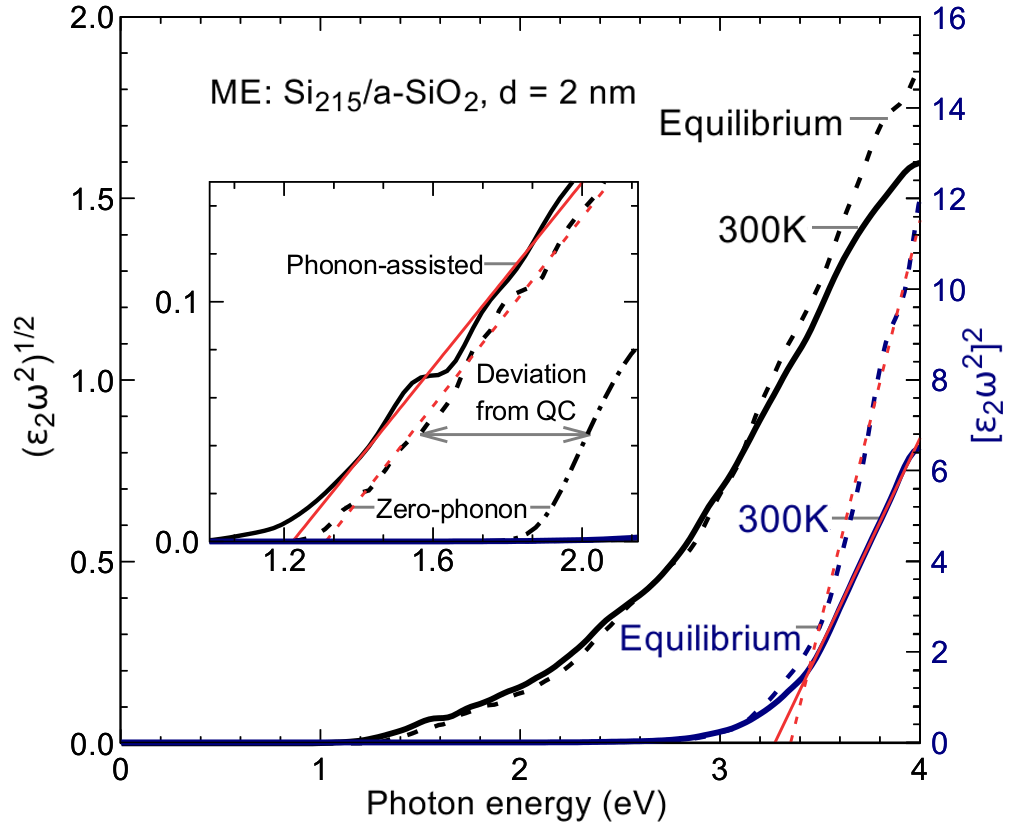}
\caption{\label{fig7} Tauc plots $[\epsilon_2 \omega^2]^{1/2}$ (black spectra) and $[\epsilon_2 \omega^2]^2$ 
(blue spectra) of the matrix-embedded SiNC, $d=2$~nm, for indirect and direct transitions.
The inset shows the Tauc plots close to the absorption onset. 
The dashed and solid lines represent spectra calculated with the nuclei at their relaxed  
equilibrium positions (zero-phonon) and for $T=300$~K (phonon-assisted), respectively. 
The thin red lines represent the corresponding linear fits in the range of photon energies  1.3 -- 2.0~eV 
and 3.4 -- 4.0~eV for $[\epsilon_2 \omega^2]^{1/2}$ and $[\epsilon_2 \omega^2]^2$, respectively.
The dashed-dotted line in the inset represents the zero-phonon spectrum of the free-standing SiNC, $d=2$~nm, 
calculated with the nuclei at their relaxed QC positions.
A Gaussian broadening of 50 meV was used for all spectra.
}
\end{figure}

Figure~\ref{fig7} shows the Tauc plots $[\epsilon_2 \omega^2]^{1/2}$ (black) and $[\epsilon_2 \omega^2]^2$ (blue)
representing indirect and direct allowed transitions in the matrix-embedded SiNC. 
We combine on the same plot the spectra calculated with the atoms at equilibrium positions (dashed curve) and 
with the atoms at thermal positions defined by the ZG displacement for $T=300$~K (solid curve). 
It can be readily seen that the Tauc plots for direct transitions reveal optical absorption for energies 
higher than the calculated band gap at $E_{\rm g}=1.435$~eV. 
These plots can be very well fitted in the energy range 3.4 -- 4.0 eV by linear regressions (red lines)  
representing the expression $[\epsilon_2 \omega^2]^{2} \propto (\hbar \omega - E^{\rm dir}_g)$, 
where $E^{\rm dir}_g$ is the first direct gap. The linear regressions cross the energy axis  
at $E^{\rm dir}_g = 3.332$~eV and $E^{\rm dir}_g(300) = 3.271$~eV which are well above than $E_{\rm g}$.
At variance with this result, Tauc plots for indirect transitions exhibit an absorption onset close to the 
band gap energy, as illustrated in the inset of Fig.~\ref{fig7}. 
In this case, linear fits to $[\epsilon_2 \omega^2]^{1/2} \propto (\hbar \omega - E_{\rm g})$ 
in the range  1.3 -- 2.0~eV give $E_{\rm g} = 1.327$~eV and $E_{\rm g}(300) = 1.209$~eV. We note that these values 
are less than the corresponding calculated band gaps by about 0.1~eV due to the 
artificial Gaussian broadening applied to our spectra.

It is evident that unlike crystalline silicon~\cite{Zacharias_2015},
the equilibrium spectrum of SiNCs is driven by zero-phonon transitions 
that behave similarly to indirect recombination channels leading to optical absorption for energies 
below $E_{\rm g}^{\rm dir}$. Furthermore, the phonon-assisted spectrum at $T=300$~K is almost a 
rigid shift of the equilibrium spectrum in the energy range 1.3 -- 2.0~eV, reflecting essentially 
the band gap difference between $E_{\rm g}(300)$ and $E_{\rm g}$. 
This latter behavior is manifest in bulk semiconductors with a direct band gap~\cite{Zacharias_2016}, 
where vertical transitions dominate. Taking these two observations together, we identify 
that optical absorption below the pseudo direct gap in nanoscale silicon structures is dominated by  
quasidirect absorption, while phonon-assisted transitions play a less important role. 

In the inset of Fig.~\ref{fig7}, we also include the spectrum of the H-terminated free-standing SiNC calculated with 
the nuclei at their relaxed QC positions (dashed-dotted curve). This shows essentially that the absorption onset of the 
matrix-embedded SiNC deviates from its QC value by more than 0.5~eV. 
Therefore, besides QC, a substantial contribution to the quasidirect transition probability 
in SiNCs originates from 
strained Si-Si bonds and oxygen surface states, as evidenced 
in Secs.~\ref{sec.Results_Role_Si-Si_strain} and~\ref{sec.Results_Size_SiNCs}. 
These effects reflect further structural modifications that enhance 
the relaxation of the $k$-conservation rule, thereby allowing for additional
zero-phonon optical transitions between the VBM and CBM.  

\section{Conclusions} \label{Sec.Conclusions}

In this manuscript we have performed a first-principles study of temperature-dependent 
optical properties of SiNCs. In a nutshell, we have demonstrated using SDM that the 
electron-phonon renormalization of the band gap of SiNCs strongly depends on 
the different passivation regimes. Starting from H-terminated free-standing SiNCs 
we have shown that the large zero-point renormalization and the band gap variation 
with temperature originates from the coupling of electrons with surface phonons. 
This surface effect is suppressed by oxidation and almost vanishes 
by inserting the nanocrystal inside the a-SiO$_2$ matrix. The present results help
to clarify the experimental measurements regarding the energy 
shifts between the temperature-dependent PL peaks of oxidized free-standing and matrix-embedded SiNCs. 
Furthermore, our data for the band gap renormalization of matrix-embedded SiNCs 
exhibits good agreement with the experiment and follows closely the Varshni behavior of bulk Si. 

Importantly, our analysis reveals that the electronic structure of SiNCs is highly 
correlated with the strain of Si-Si bonds. 
In fact, in the case of matrix-embedded SiNCs, we have demonstrated a strong linear dependence between 
the band gap and the bond length variance induced by atomic vibrations. 
This result suggests the inherent relationship between electron-phonon coupling and thermally 
averaged structural perturbations that can be explored with non-perturbative approaches, like SDM. 

Beyond studying the temperature dependence of the band gap, 
we investigate the effect of phonon-assisted electronic transitions 
on the optical spectra of SiNCs. At variance with bulk Si, 
the spectrum close to the absorption onset of SiNCs is dominated by zero-phonon 
quasidirect transitions, while phonon-assisted recombination is less important.
We also clarify that the origin of quasidirect transition probability in SiNCs is not only 
due to quantum confinement, but also due to strained Si-Si bonds (primarily)
and oxygen-related surface states (secondarily).

Finally, it should be possible to extend our present work to
explore how the formation of oxygen bridge bonds, or other oxidation combinations 
can explicitly affect electron-phonon coupling in SiNCs. 
Furthermore, we expect that the calculation of temperature-dependent band gaps 
of other important nanostructures, such as Ge, SiGe and TiO$_2$ passivated nanocrystals
should be within reach. Our study can also  
be upgraded with calculations of full photoluminescence spectra accounting 
for excitonic effects and exciton-phonon coupling~\cite{Paleari_2019} via the combination 
of Bethe-Salpeter approach and SDM.

\acknowledgments

MZ and PC thank Christos Mathioudakis and George Hadjisavvas for help 
in generating the structures of SiNCs embedded in amorphous SiO$_2$.
This work was supported by the Strategic Infrastructure Project NEW
INFRASTRUCTURE / $\Sigma$TPATH / 0308 / 04, which is
co-funded by the European Regional Development Fund, the European
Social Fund, the Cohesion Fund, and the Research Promotion Foundation
of the Republic of Cyprus. The calculations were supported by the Cy-Tera Project 
(NEW INFRASTRUCTURE / $\Sigma$TPATH / 0308 / 31), which is co-funded by the European Regional Development Fund and 
the Republic of Cyprus through the Research Promotion Foundation.
 All electronic structure calculations performed 
in this study are available on the NOMAD repository: http://dx.doi.org/10.17172/NOMAD/2020.04.30-1

\bibliography{references}{} 
\end{document}